\documentclass[sigconf,screen]{acmart}
\usepackage[utf8]{inputenc}
\usepackage{microtype}
\usepackage{graphicx}
\usepackage{makecell}
\usepackage{array}
\usepackage{subfig}
\usepackage{standalone}
\usepackage{tikz}
\usetikzlibrary{arrows,automata,arrows.meta}
\tikzstyle{mybox} = [draw=black, thick,rectangle, inner sep=10pt]
\tikzset{vertex style/.style={
fill=#1!70,
text=black,
rounded rectangle,
minimum width=2cm,
minimum height=0.75cm,
font=\normalsize,
outer sep=3pt, %
},
}

\newcommand\blfootnote[1]{%
  \begingroup
  \renewcommand\thefootnote{}\footnote{#1}%
  \addtocounter{footnote}{-1}%
  \endgroup
}

\title{AlgebraicSystems: Compositional Verification for~Autonomous~System~Design}
\author{Georgios Bakirtzis}
\affiliation{%
  \institution{The University of Texas at Austin}
}
\email{bakirtzis@utexas.edu}

\author{Ufuk Topcu}
\affiliation{%
  \institution{The University of Texas at Austin}
 }
\email{utopcu@utexas.edu}

\renewcommand\footnotetextcopyrightpermission[1]{}

\begin{document}
\begin{abstract}
Autonomous systems require the management of several model views to assure properties such as safety and security among others. A crucial issue in autonomous systems design assurance is the notion of emergent behavior; we cannot use their parts in isolation to examine their overall behavior or performance. Compositional verification attempts to combat emergence by implementing model transformation as structure-preserving maps between model views. AlgebraicDynamics relies on categorical semantics to draw relationships between algebras and model views. We propose AlgebraicSystems, a conglomeration of algebraic methods to assign semantics and categorical primitives to give computational meaning to relationships between models so that the formalisms and resulting tools are interoperable through vertical and horizontal composition.
\end{abstract}
\maketitle
\pagestyle{plain}
\section{Motivation}
\begin{figure}[!t]
    \centering
    \includegraphics[width = .8\linewidth]{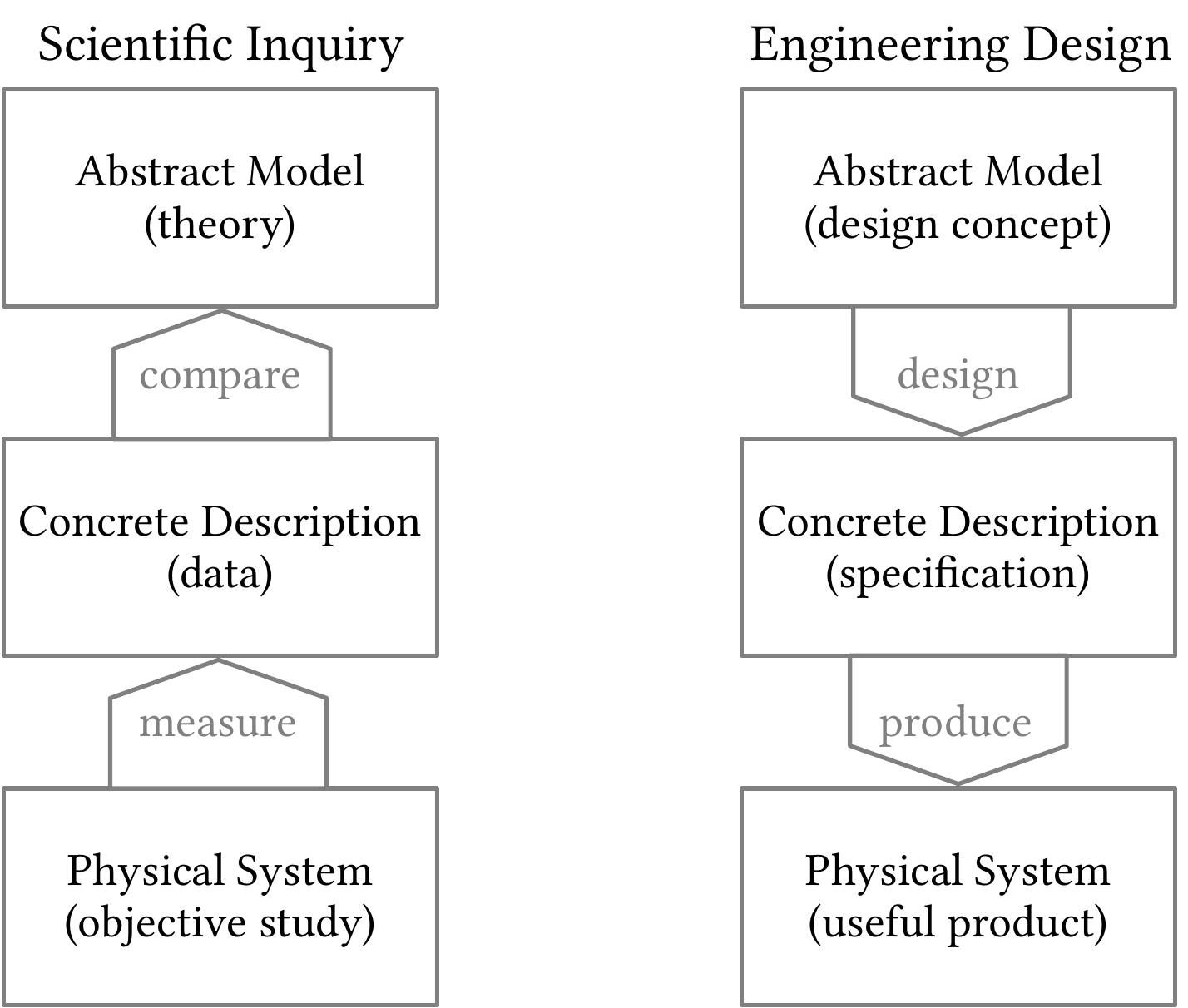}
    \caption{The value of an engineered system is how well it matches its associated models (adapted from Drextel~\cite{drexler:2013}).}
    \label{fig:engineering}
    \vspace{-2em}
\end{figure}

Ensuring\blfootnote{This research is based upon work supported by NASA 80NSSC21M0071.} that autonomous systems will behave as expected based on their requirements is often achieved by modeling. Model-based design is different from science (Figure~\ref{fig:engineering}). In science, we deal with systems that we have no control over, and we attempt to create formal models of, for example, how things move based on assumptions about the environment, inching with different paradigms closer to reality. Instead, in engineering, we have the benefit of amalgamating systems \emph{from} models; that is, the value of our realized systems is how well we can conform them to our understanding captured in models~\cite{lee:2021}. Here too, we deal with different paradigms. In autonomous systems, the different paradigms can be viewed explicitly or implicitly as residing within distinct algebras.

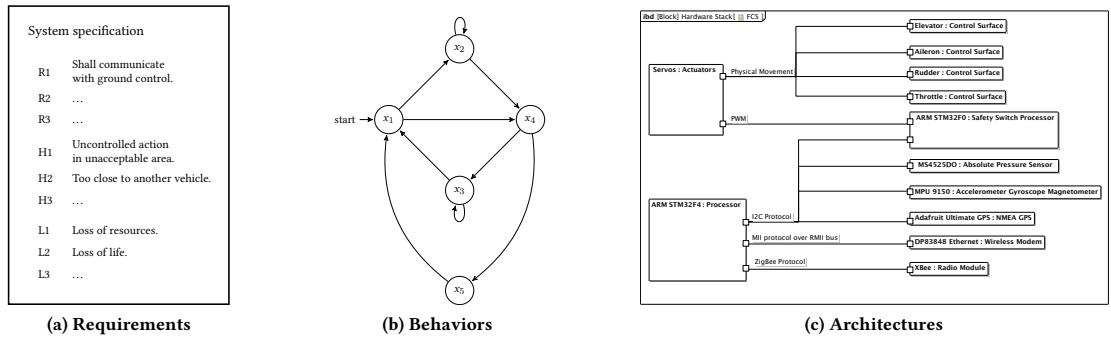
\begin{figure*}[!t]
    \centering
    \subfloat[Requirements]{\scalebox{.75}{
    \begin{tikzpicture}
\node [mybox] (box){%
    \begin{minipage}{0.18\textwidth}
        {\footnotesize System specification}\\[1em]
        \begin{tabular}{>{\scriptsize} l >{\scriptsize} l}
        R1 & \makecell[l]{Shall communicate\\ with ground control.}\\
        R2 & \dots \\
        R3 & \dots \\[.5em]
        H1 &  \makecell[l]{Uncontrolled action\\ in unacceptable area.}\\
        H2 &  Too close to another vehicle.\\
        H3 &  \dots \\[.5em]
        L1 &  Loss of resources.\\
        L2 &  Loss of life.\\
        L3 &  \dots \\
        \end{tabular}
    \end{minipage}
};
\end{tikzpicture}}
}\qquad\qquad
  \subfloat[Behaviors]{\begin{tikzpicture}[->,>=stealth',shorten >=1pt,auto,node distance=2.8cm,
                    semithick]
  \tikzstyle{every state}=[text=black]

  \node[initial,state] (A)                    {$x_1$};
  \node[state]         (B) [above right of=A] {$x_2$};
  \node[state]         (D) [below right of=A] {$x_3$};
  \node[state]         (C) [below right of=B] {$x_4$};
  \node[state]         (E) [below of=D]       {$x_5$};

  \path (A) edge              node {} (B)
            edge              node {} (C)
        (B) edge [loop above] node {} (B)
            edge              node {} (C)
        (C) edge              node {} (D)
            edge [bend left]  node {} (E)
        (D) edge [loop below] node {} (D)
            edge              node {} (A)
        (E) edge [bend left]  node {} (A);
\end{tikzpicture}}
\qquad\qquad
\subfloat[Architectures]{\includegraphics[width=.35\textwidth]{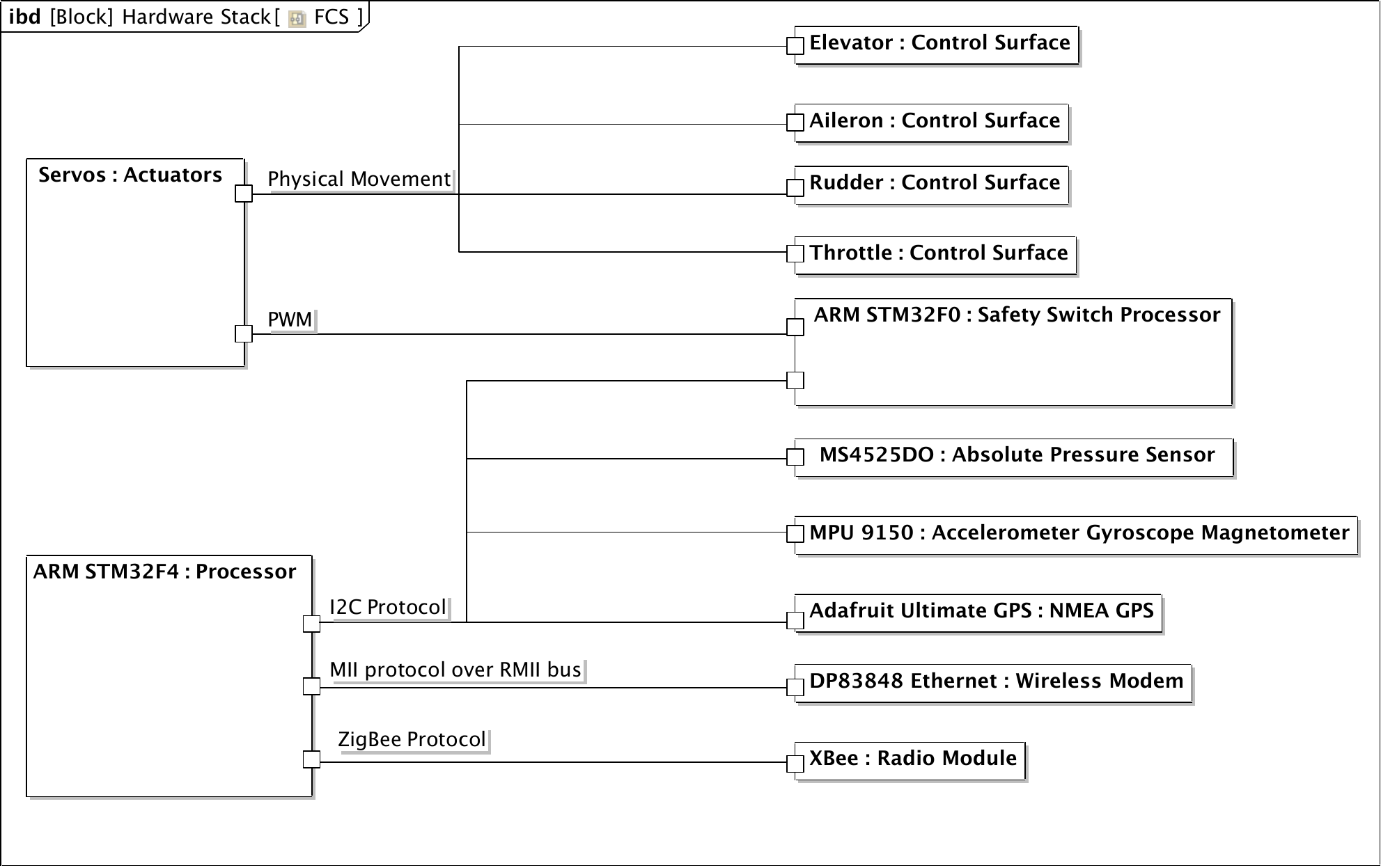}}
\vspace{-1em}
    \caption{There is a semantic gap between types of models needed to assure safe and secure behavior of deployed systems.}
    \label{fig:semantics}
\end{figure*}

Viewing models as algebras allows us to reason compositionally between them. However, most engineering work in compositionality centers around a particular formalism. For example, hybrid systems \cite{nejati:2021} and timed automata \cite{bouyer:1999} model behavior, linear temporal logic specifications \cite{alur:2018} and contracts \cite{ghasemi:2020} model requirements. We focus on the composition of individual algebras rather than their relation. Call this horizontal-type composition; within one mathematical model, we compose the same types of models to produce larger ones. Horizontal composition is generally an accepted line of work within one field. Still, a perhaps more interesting rule would be vertical composition, which would relate or otherwise be able to enforce a hierarchy among multiple such formalisms (Figure~\ref{fig:semantics}). One way of achieving vertical composition is by using tools from category theory.

Compositional verification refers to the rules that connect individual parts to construct a whole in a way that the behavior determines the whole. This is to say that compositionality is about \emph{refinement} and \emph{abstraction}~\cite{hedges:2016}. Refinement is about augmenting or recovering more information about a particular component. But abstraction is equally important in designing increasingly complex autonomous systems, which require us to put components together that form larger systems via black-boxing every component in such a way that, when composed, defines the desired system's behavior.

One way to relate the algebras of requirement, behaviors, and architectures is to carve a common research trajectory with applied category theory. In engineering, category theory can give precise meaning to the transformations associated with remembering and forgetting between model views. To date, these advances are predominantly theoretical. To be fruitful in the program of compositional systems theory~\cite{bakirtzis:2021,bakirtzis:2021d}, we must develop a computational interpretation of algebras and their associated horizontal and vertical composition rules. The eventual user of these tools should not have to be an expert in category theory but rather be able to leverage (1) warnings raised when composition fails and (2) have interoperability between theories, tools, and consequently analyses and synthesis techniques. AlgebraicSystems is an envisioned comprehensive program that implements these ideas in the high-performance Julia programming language~\cite{bezanson:2017}.

Interoperability between models and tools outputs assurance cases for properties we care about, such as safety and security, through compositional verification~\cite{bakirtzis:2021b}. By designing systems compositionally, we precisely address the lack of interoperability between formalisms, an open problem~\cite{lukcuck:2019}, within AlgebraicSystems.

\section{Compositional Verification}

A combination of models is often used to assure metrics such as safety and security and dynamics and control (Figure~\ref{fig:semantics}). The categorical formalization of composition gives rise to the unification of (a subset) of requirements, system behaviors, and system architectures in a traceable manner. The categorical primitive of functoriality---structure-preserving maps between categories---gives concrete meaning to abstraction and refinement in cyber-physical system models, which can assist with the specification (and eventually validation) of increasingly complex systems.

Several mathematical domains, such as graph theory, can implement compositional verification. In compositional systems theory and AlgebraicSystems, verification is not about prediction but rather abstracting and refining structure---organizing different models within categories for syntax and assigning algebras for semantics. Category theory is one context where the meaning of composition is formal and refers to something specific, namely the partial operation on morphisms of a category. We formalize the notion of composite systems in this work using the systems-as-algebras framework in the wiring diagram category.

For example, a controls model can have a syntactic flavor within the wiring diagram category, where we precisely construct formal interfaces between boxes~\cite{bakirtzis:2020,bakirtzis:2021a}. At the moment, there is no behavior, just an architectural arrangement of parts. It is then the job of the controls algebra to assign meaning to each of the boxes. The computation of the dynamic behavior of the control system is then the horizontal composition of boxes that are inhabited by controls algebras (semantics) and how those semantics give rise to the whole given the arrangement between those boxes with wires (syntax). 

There are several models and assurance methods we would like to relate compositionally for autonomous systems. Types of models and assurance methods for autonomous systems that are not currently related or have compositional verification between them include models of Markov decision processes~\cite{shiebler:2021}, control synthesis~\cite{tabuada:2006}, contracts~\cite{zenodo:2022}, and shielding~\cite{konighofer:2017}. AlgebraicSystems is to be the conglomeration of these models and methods.

AlgebraicSystems gives formal and computational meaning to relationships between formalisms, domain-specific models, and assurance methods. We can treat syntactic elements as categories and semantics as algebras by implementing compositional verification using category theory. The categorical interpretation of system theory engenders an understanding of model-based design as examining how formalisms are \emph{related} to each other rather than how individual formalisms model systems in isolation.
\bibliographystyle{ACM-Reference-Format}
\bibliography{manuscript}
\end{document}